\documentclass[12pt]{article}
\usepackage[dvips]{graphicx}
\usepackage[hang]{subfigure}

\newcommand{\bfi}{\begin{figure}}
\newcommand{\efi}{\end{figure}}
\newcommand{\subf}{\subfigure}
\newcommand{\bmi}{\begin{minipage}}
\newcommand{\emi}{\end{minipage}}

\title{A Circular Statistical Method for Extracting Rotation Measures}
\date{}
\author{S. Sarala and Pankaj Jain \\
Physics Department, IIT, Kanpur 208 016, India }

\begin{document}

\maketitle

\begin{abstract}
We propose a new method for the extraction of Rotation Measure from
spectral polarization data. The method is based on maximum 
likelihood analysis and takes into account the circular nature
of the polarization data.  
 The method is unbiased
and statistically more efficient than the standard $\chi^2$ procedure.  
We also find that 
the method is computationally much faster than the standard $\chi^2$
procedure if the number of data points are very large. 
\end{abstract}

\section {Introduction}
Polarizations of radio waves from cosmologically distant
sources undergo Faraday rotation
upon propagation through galactic magnetic fields.
This effect provides very useful information about the
magnetic fields in our galaxy as well as in the host
galaxy (Vall\'{e}e 1997; Zeldovich 83). 
The amount of rotation is proportional to the
magnetic field component parallel to the direction
of propagation of the wave and to the square of the
wavelength $\lambda$. The observed orientation of
the linearly polarized component of the electromagnetic 
wave $\theta (\lambda^2 )$ can therefore be written as,
 
\begin{eqnarray*}
\theta (\lambda^2 ) = \theta_0 + \left( RM
\right)\lambda^2 
\end{eqnarray*}
where the slope, called Rotation Measure ($RM$), depends linearly
on the parallel component of the magnetic field and $\theta_0$ is
the intercept, also called the 
intrinsic position angle of polarization, $IPA$.

$RM$ and $IPA$ can be determined by the standard $\chi^2$ procedure. 
 The extraction of $RM$ is ambiguous since 
the observed polarization is defined only  
upto additions of $n\pi$, where $n$ is an integer 
(see for example, Simard-Normandin et al. 1981; Zeldovich et al. 1983; Broten et al. 1988).
Therefore one finds many fits depending on the choice of $n \pi$
for the polarizations at different wavelengths and it is necessary to find
the $RM$ which gives the absolute minimum value of $\chi^2$.
However, for small data sets choosing absolute minima is not reliable 
since it is often possible
to make $\chi^2$ arbitrarily small by making $RM$ sufficiently large.
It is therefore necessary to limit the range of $RM$ while searching for
the minimum $\chi^2$. 

Another problem that also arises in many cases
is that the observed polarization does not actually show a linear
dependence on $\lambda^2$ (Vall\'{e}e 1980). 
This can arise if different regions
of the source galaxy dominate at different wavelengths. It has,
therefore been recommended that one should concentrate on a narrow
region of the wavelength where the straight line may be a good fit
to data. A methodology for the selection of limited wavelength region 
based on a unique physical regime has been suggested by Vall\'{e}e (1980)
and used by  Broten, MacLeod \& Vall\'{e}e (1988) to compile a catalogue of
unambiguous $RM$. 

In the present paper we propose an alternative method for the extraction
of $RM$ and $IPA$ from data. This procedure  
 is based on maximum 
likelihood for distributions defined to be invariant under angular 
coordinate changes.  The  von Mises (vM) distribution serves as a prototype for 
the statistical fluctuations for circular data. It is given by,
(Mardia, 1972; Batschelet, 1981; Fisher, 1993)
\begin{equation}
f_{\rm vM}(\theta) = {\exp \left[ \kappa\cos2(\theta-\overline \theta)
\right] \over \pi I_0(\kappa)}
\label{von}
\end{equation} 
where $\kappa$ is a parameter which measures the concentration of the
population and $\overline \theta$ is the mean angle. The factor
2 has been inserted since the polarization angle is ambiguous by $n\pi$
and not $2n\pi$ (Born and Wolf, 1980; for a detailed discussion on this 
point see Ralston \& Jain, 1999). 
The von Mises distribution for circular data is in many ways the analoque of
normal distribution for linear data 
 (Mardia, 1972; Batschelet, 1981; Fisher, 1993).   
For example, a two dimensional normal distribution $g(x,y)$ restricted
to the circle $x^2 + y^2 = 1$ is equivalent to the von Mises distribution.
For large $\kappa$ the von Mises distribution is strongly peaked
at $\theta=\overline \theta$. Keeping only the leading order term in
$\theta-\overline \theta$ we find that it reduces to
the normal distribution in this limit. 
The von Mises distribution has found applications in many physical 
examples involving circular data. 
For example, it has been
found (Kendall and Young, 1984; Jain and Ralston, 1999) that the  
variable $\beta=\theta_0-\psi$, where $\theta_0$ is the
IPA and $\psi$ is the orientation
axis of the radio galaxy, is well described by this distribution. 

We can obtain the distribution of the polarization angle by
making the reasonable assumption
that the electric field components follow a normal distribution
(Brosseau, 1998),
$$p(E_1,R_2) = {det({\bf R})\over (\pi^2)}\exp\left(-\sum_{i,j=1}^2
E_i^*R_{ij} E_j\right) $$
We work in the circular polarization basis and $E_1,E_2$ here 
refer to the right and left circularly polarized components of
the electric field. $R_{ij}$ are the parameters of the distribution.
We make a change of variables such that $E_j=a_j\exp(i\theta_j)$.
We are interested in the distribution of the linear polarization angle
$\theta$, which in the circular polarization basis is equal to
$(\theta_1-\theta_2)/2$. After integrating over the rest of the variables
we obtain the MacDonald-Bunimovitch (MB) 
distribution $p_{MB}(\theta)$ (Brosseau 1998, MacDonald 1949, Bunimovitch 1949),
\begin{equation}
p_{MB}(\theta) = {(1-\xi^2)\over \pi (1-\mu^2)^{3/2}}\left[\mu\sin^{-1}\mu
+{\pi\over 2}\mu + (1-\mu^2)^{1/2}\right]
\label{sharp}
\end{equation}
where $\mu=\xi\cos(2\theta-2\overline\theta)$. Here $\xi^2=1-|R_{12}|^2/R_{11}
R_{22}$ $(0\le \xi \le 1)$ and $\overline\theta$ 
are the parameters of the distribution which
measure the concentration and mean value of the population respectively. 

\begin{figure}
\hskip 1.0in
\includegraphics[scale=1.0]{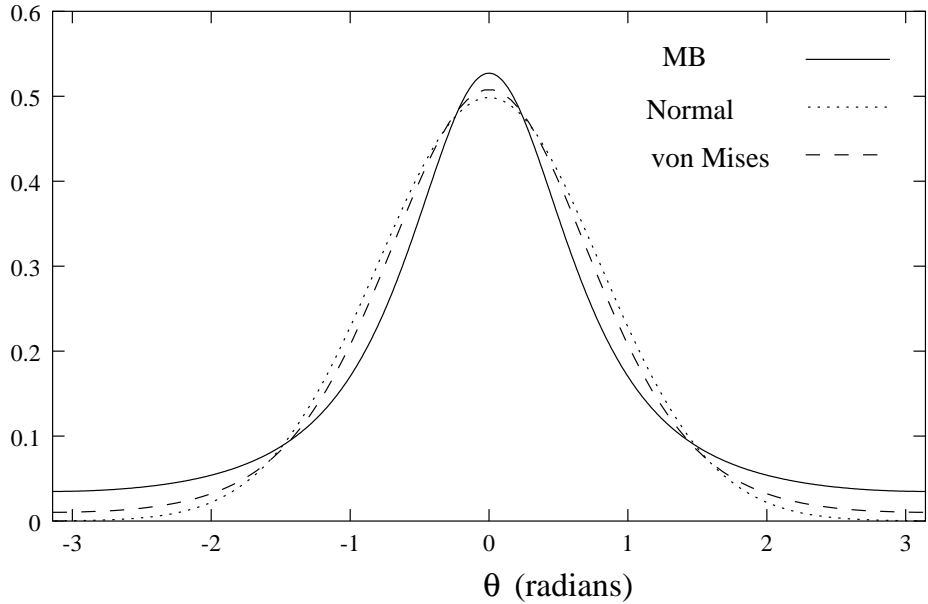}
\caption{Comparison of the normal distribution ($\sigma=0.8$) 
with von Mises and
the MB distribution, Eq. \ref{sharp}. The 
parameters of the von Mises and MB were determined by 
requiring that the least square difference between them and the normal
distribution is minimum.} 
\label{distcomp}
\end{figure}

The von Mises as well as the normal distribution reduce to the $p_{MB}(\theta)$ 
in the limit when the population is strongly concentrated near the
mean value, i.e. in the limit when the parameter $\xi\rightarrow 1$.
However if the distribution is not too narrow these distributions 
differ significantly as shown in fig. \ref{distcomp}. Furthermore the
von Mises distribution and $p_{MB}(\theta)$ are invariant under angular
coordinate translations which is not the case for normal distribution. 
The MB distribution turns out to be more sharply peaked
in comparison to von Mises. 
It is interesting that the distribution of 
$\beta$, the difference between the polarization angle and galaxy orientation
angle, was empirically found to be better described by a more sharply peaked
function in comparison to von Mises (Jain and Ralston, 1999).

In our likelihood analysis we find it convenient to first determine 
the rotation measure and IPA using the von Mises distribution. 
This is done by maximizing the log likelihood function numerically.
This extremization 
turns out to be relatively easy since it is possible
to analytically eliminate one of the variables. It is then only necessary to
perform a one dimensional extremization numerically.   
The rotation
measure and IPA determined by this procedure then serves as an initial
guess for the analysis using the more complex MB distribution, 
Eq. \ref{sharp}, for which we have to maximize the log likelihood
function over both the variables numerically.
Since the polarization 
angle is expected to follow this distribution 
our procedure 
will yield an unbiased and statistically efficient measure 
(see for example, Freund, 1992; Cramer, 1958) of the $RM$
in comparison to the standard method of $\chi^2$ minimization.
The $\chi^2$ procedure is based on the gaussian distribution which
is a not invariant under angular coordinate changes and hence cannot
provide an optimal description for angular data.
We point out that the standard theorems, which assure that
the $\chi^2$ estimate is unbiased and of minimal variance, are
not applicable to circular data. 
 
Besides the fact that the maximum likelihood procedure is more efficient
we find that it is also computationally less intensive in the case that
the number of data points are large. In the case of standard $\chi^2$
procedure a fit has to be made for all possible choices of the 
polarization angle $\theta$ obtained after addition by $\pm n \pi$
$(n = 0, \pm 1, \pm 2, \cdots )$.  
With the increase in the number of data points  in the chosen wavelength
range the total number of combinations increase very rapidly requiring 
very long computation time to exhaust all the possible choices. 
However, large number of data points are preferable in order to 
reduce the ambiguities
in the extraction of $RM$ and $IPA$. We find that the computational
time for the current procedure increases very slowly with the number of
data points. This is because for the case of von Mises distribution
it only requires a search over one parameter. 
By explicit calculations we find that the program converges to the
absolute 
minima relatively fast without requiring an unduly small search interval.
This solution is then
used as an input for the more realistic
MB distribution (Eq. \ref{sharp}), which also turns out to converge very
fast to the true solution.   
Hence the procedure can be very useful in extracting unambiguous 
$RM$ from data.

We apply our procedure to the spectral data given in the catalogue of
linear polarization by Tabara \& Inoue (1980). The catalogue 
contains position angle of polarization as a function of $\lambda^2$ 
for 1510 sources. However, for many of these sources there exist only 
one or two data points and hence the extraction of $RM$ is not possible.
For the remaining 704 sources $RM$ is listed in the catalogue.
This is obtained by limiting the range of $RM$ search
to less than 200 $rad/m^2$ except on the galactic plane ($\cot |b| <2$).
For $\cot |b|>2$ the range is extended to 100 $\cot |b| rad/m^2$.
For three of these 704 sources, data is listed in the catalogue at only two
wavelengths. We ignore these three sources for our analysis and apply
our procedure for the remaining 701 sources. 

\section {Likelihood Method for Extracting RM }

Our procedure is based on the maximum likelihood analysis using 
a distribution for polarization angle obtained under the assumption
that the electric field vector is normally distributed.
We find it useful in actual computations to first use a simpler
form for the distribution function which is
invariant under angular coordinate transformations. 
We assume this to be the  
von Mises distribution given in equation (\ref{von}).
Since the mean angle $\overline \theta=RM \lambda^2 + \theta_0$, the
distribution can be written as,

\begin{equation}
f(\theta;RM,\theta_0) = {\exp \left[\kappa\cos 2(\theta-RM\lambda^2-\theta_0)
\right] \over \pi I_0(\kappa)}
\label{distri}
\end{equation} 
The best fit parameters $RM,\theta_0$ can be obtained by maximizing
the log likelihood of the spectral polarization data $\theta(\lambda^2)$.
Let ${\theta_i}$ and ${\lambda_i}$ denote the measured polarization
angle and the corresponding wavelength respectively. The log likelihood
of the data is then given by,
\begin{equation}
\ln{\cal L} = \sum_{i=1}^N \kappa_i \cos 2(\theta_i-RM\lambda_i^2-\theta_0)
- N\ln \pi - \sum_{i=1}^N \ln I_0(\kappa_i)
\label{logl}
\end{equation}
where $N$ is the total number of data points for a given source. 
We can estimate the parameter $\kappa_i$ from the error in the measurement
of the polarization angle. For a circular variable $2\theta_i$ the measure of
concentration (Batschelet, 1981) is given by the length of the mean vector 
\begin{equation}
 r_i = \left(1 \over M \right) \sum_{j=1}^M \cos 2\left( \theta_{ij} - 
\overline \theta_i \right)\label{ri}
\end{equation} 
where $\theta_{ij}$ refers to a particular measurement of $\theta_i$
and $M$ are the total number of polarization 
measurements taken at a particular
value of $\lambda_i$.
For von Mises distribution 
\begin{equation}
 r_i = {I_1 (\kappa_i)\over{I_0(\kappa_i)} }
\label{ri1}
\end{equation}
where $I_\nu$ is the modified Bessel function of order $\nu$.
Assuming that the error in the measurement of $\theta_i$ is small, we expect
 $\kappa_i$ to be large and $r_i \approx 1$. Expanding the Bessel function for
large value of $\kappa_i$ we find that 
$$
\kappa_i = {1\over{2 (1-r_i)}}$$
If the assumption that the errors are not small is not valid then 
it is better to solve Eq. \ref{ri1} numerically. 
Since the error in the measurement of $\theta_i$ is $\Delta\theta_i$ we find
that $r_i = \cos(2\Delta\theta_i)$.
Therefore
$$ \kappa_i = {1\over{ 2\left( 1 - \cos(2 \Delta \theta_i)\right) }}$$
Substituting this in the equation (\ref{logl}) and maximising the log
likelihood with respect to $RM$ and $\theta_0$ gives the estimate
of these parameters. This turns out to be equivalent to minimizing

\begin{equation}
\chi^2_{cir} = \sum_{i=1}^N { {1 - \cos 2 \left[ \theta_i - \theta_0 - \left(RM
\right) \lambda_i^2 \right] \over
 {1 - \cos 2 \Delta \theta_i} }} 
\label{chi_cir}
\end{equation}
which is analogous to the standard $\chi^2$ used in the case of linear
variables. For a large $\kappa_i$ we find that $\chi^2_{cir}$  
follows 
the $\chi^2$ distribution with $N$ degrees of freedom, the same as that
followed by the standard linear $\chi^2$. We find by explicit calculations
that the $\chi^2_{cir}$ is close to the usual $\chi^2$ for most of
the sources.
Minimising with respect to $\theta_0$ gives 
\begin{eqnarray*}
\tan 2 \theta_0 = 
\frac
{\sum_i  \sin B_i/ \left( 1 - \cos 2 \Delta \theta_i \right)}
{ \sum_i { \cos B_i/ \left( 1 - \cos 2 \Delta \theta_i \right)}
}
\end{eqnarray*}
with $B_i = 2 [\theta_i - \left(RM
\right) \lambda_i^2 ]$.
Substituting this into equation (\ref{chi_cir}) we can find
the minimum of $\chi^2_{cir}$ by searching numerically over a single parameter
$RM$. 

For the case of more realistic MB distribution, Eq. 
\ref{sharp}, we again maximize the log likelihood,
$$\ln {\cal L} =  \sum_{i=1}^N \ln[p_{MB}(\theta_i)] $$ 
as a function of the parameters $RM$ and $\theta_0$ in exact analogy
as was done for the von Mises distribution. The main difference
in the present case is that it is not possible to eliminate 
$\theta_0$ analytically. Hence this distribution requires a two
dimensional extremization. We use the results obtained with von Mises
distribution as an initial guess for the present case. The two
dimensional extremization uses the program 'powell' (Press {\it et al} 1986) 
which gives very
fast convergence. We point out that the parameter $\xi$ in the present
case can be estimated for each polarization measurement from the
corresponding error in that measurement. Since the errors are quoted
assuming gaussian statistics, we find the parameter $\xi$ such that 
the corresponding MB distribution is closest
to the normal
distribution 
for the measured polarization angle and its error. By being
closest we mean that   
the square difference between the two distributions is minimum. 
By explicit calculation we find that for small values of the standard
deviation parameter $\sigma$ used in normal distribution, 
$1-\xi^2\approx 1.3 \sigma^2$. In order to obtain an estimate of
the goodness of 
fit by MB distribution we define the log likelihood difference 
\begin{equation}
\Delta(\ln{\cal L}) =\sum_{i=1}^N \left[\ln[p_{MB}(0)]-  \ln[p_{MB}
(\theta_i)]\right]\ .
\label{LikDiff}
\end{equation}
By making a Taylor expansion around $\theta_i=0$ and keeping terms upto
$\theta_i^2$ this reduces to the standard $\chi^2$ upto an overall constant
equal to $6/1.3$. The $1.3$ in the denominator arises due to factor
of $1.3$ between $\sigma$ corresponding to gaussian statistics and
$\xi$ which appears in MB distribution. We find by explicit calculations
that $\Delta(\ln{\cal L})$ is proportional to 
the standard $\chi^2$
only for very small errors in the polarization
angle. For large value of $\chi^2$ it does not have a linear dependence
on $\chi^2$. 
In giving results for the MB distribution we quote this
log likelihood difference. The $\chi^2/dof=1$ for gaussian statistics,
which represents the cutoff point for what is considered as a good fit,
correspond in the present case to $\Delta(\ln{\cal L})/dof \approx  
2$. For $\chi^2/dof>1$ we find that the
$\Delta(\ln{\cal L})/dof$ increases very
slowly with increase in $\chi^2$. 

\bfi
\subf{
\bmi[t]{3in}
\includegraphics[scale=0.55]{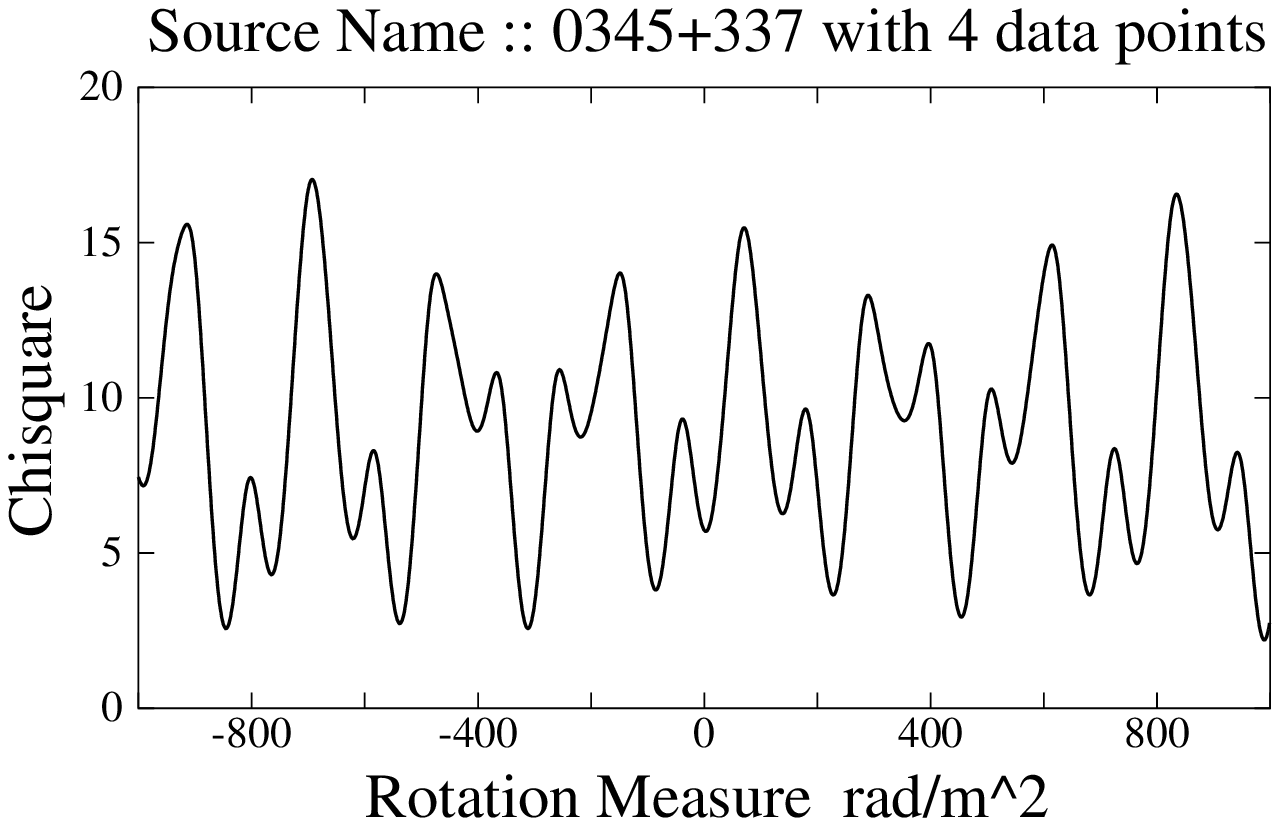}
\emi
}
\subf{
\bmi[t]{3in}
\includegraphics[scale=.55]{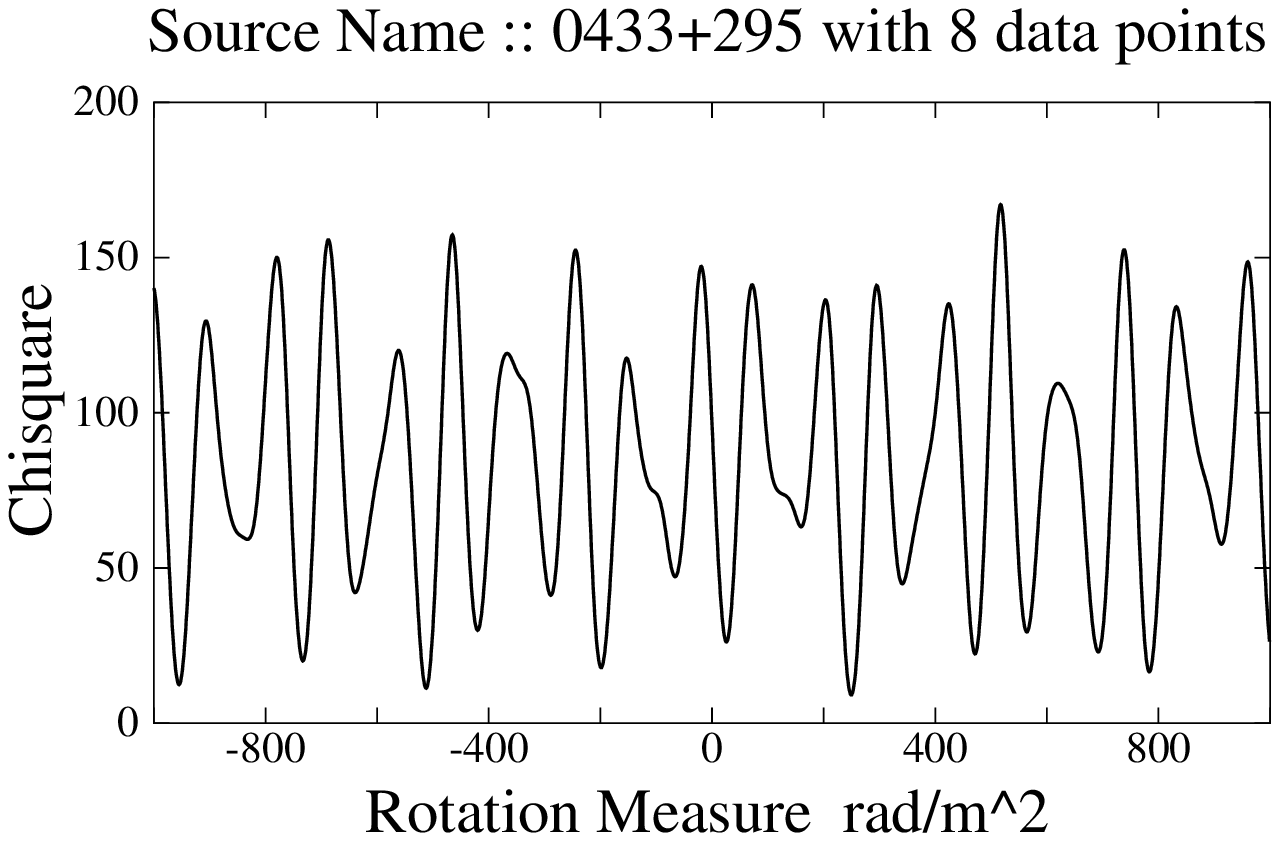}
\emi
}
\subf{
\bmi[b]{3in}
\includegraphics[scale=.55]{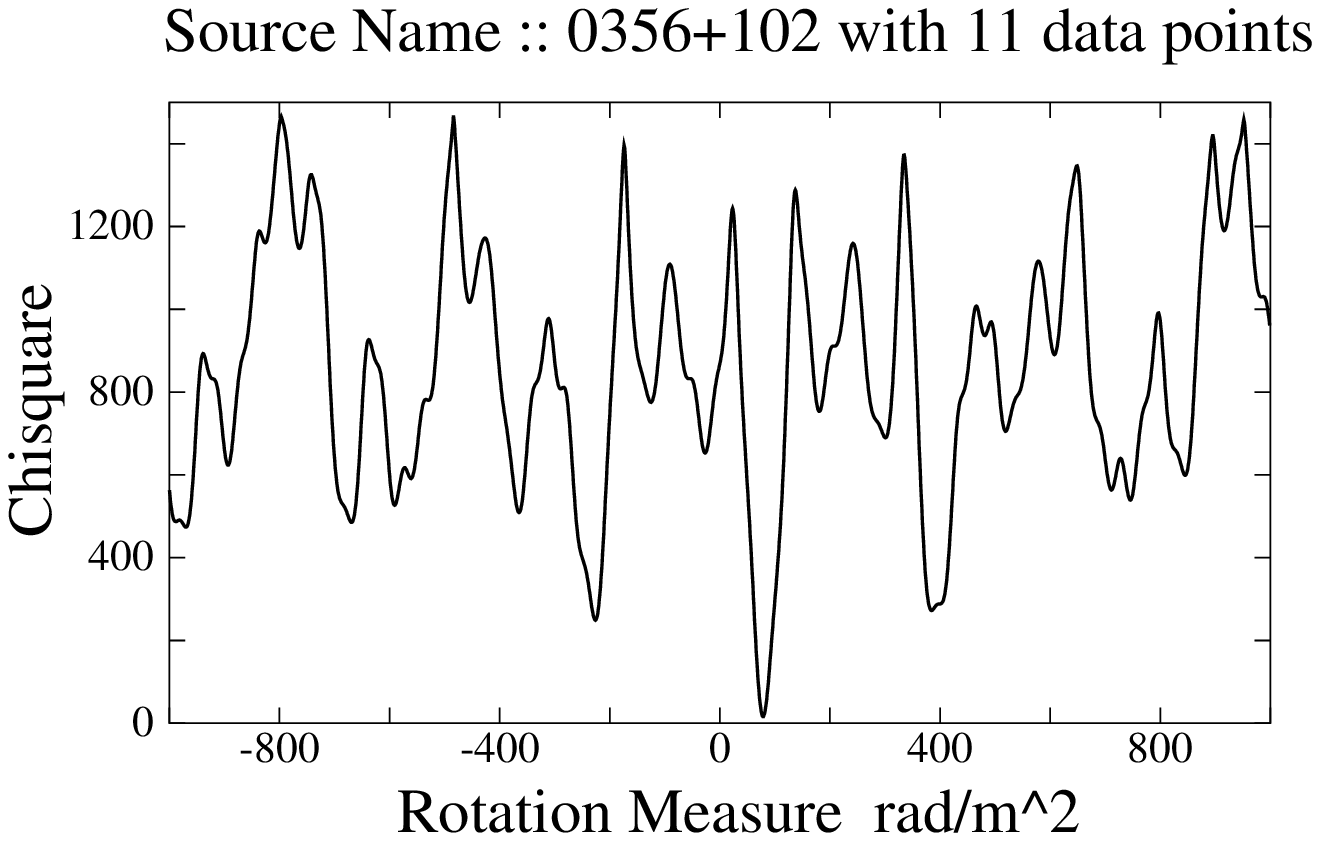}
\emi
}
\subf{
\bmi[b]{3in}
\includegraphics[scale=.55]{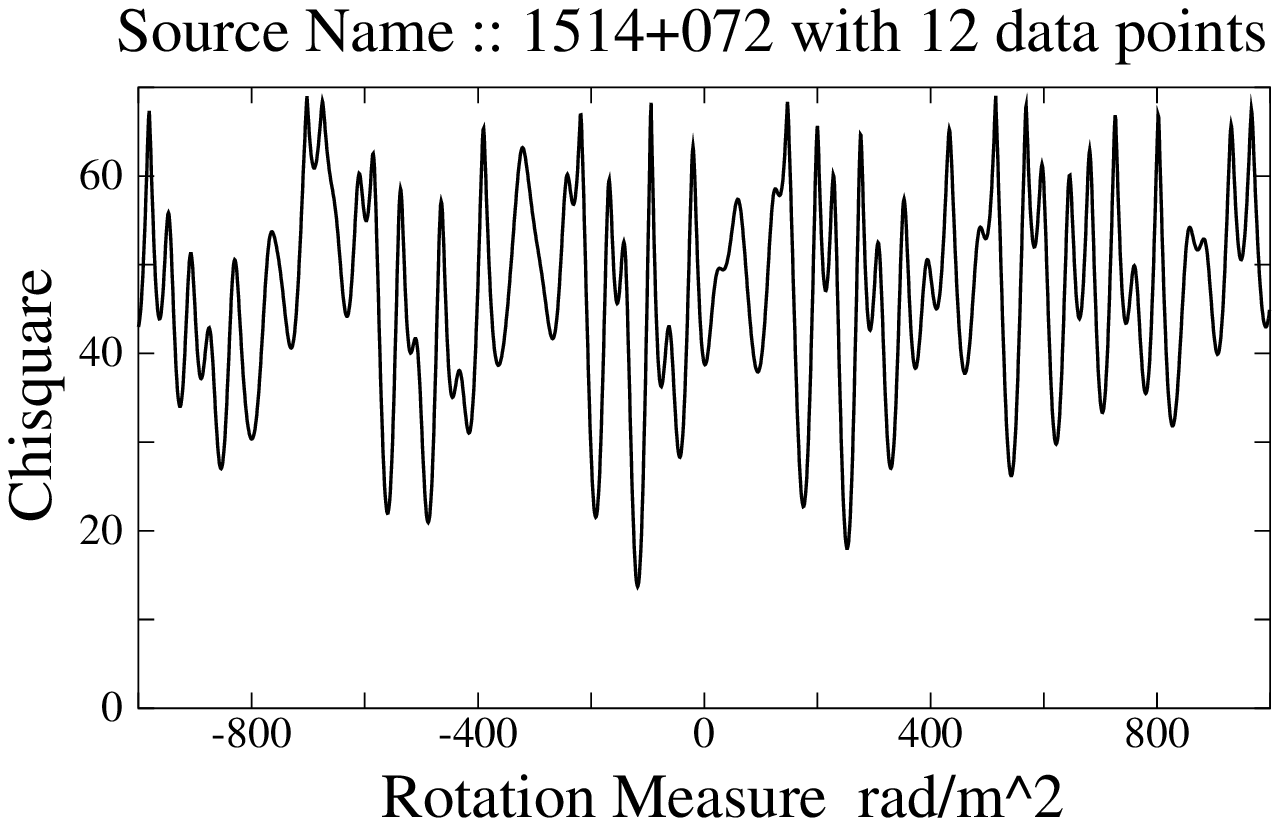}
\emi
}
\subf{
\bmi[t]{3in}
\includegraphics[scale=.55]{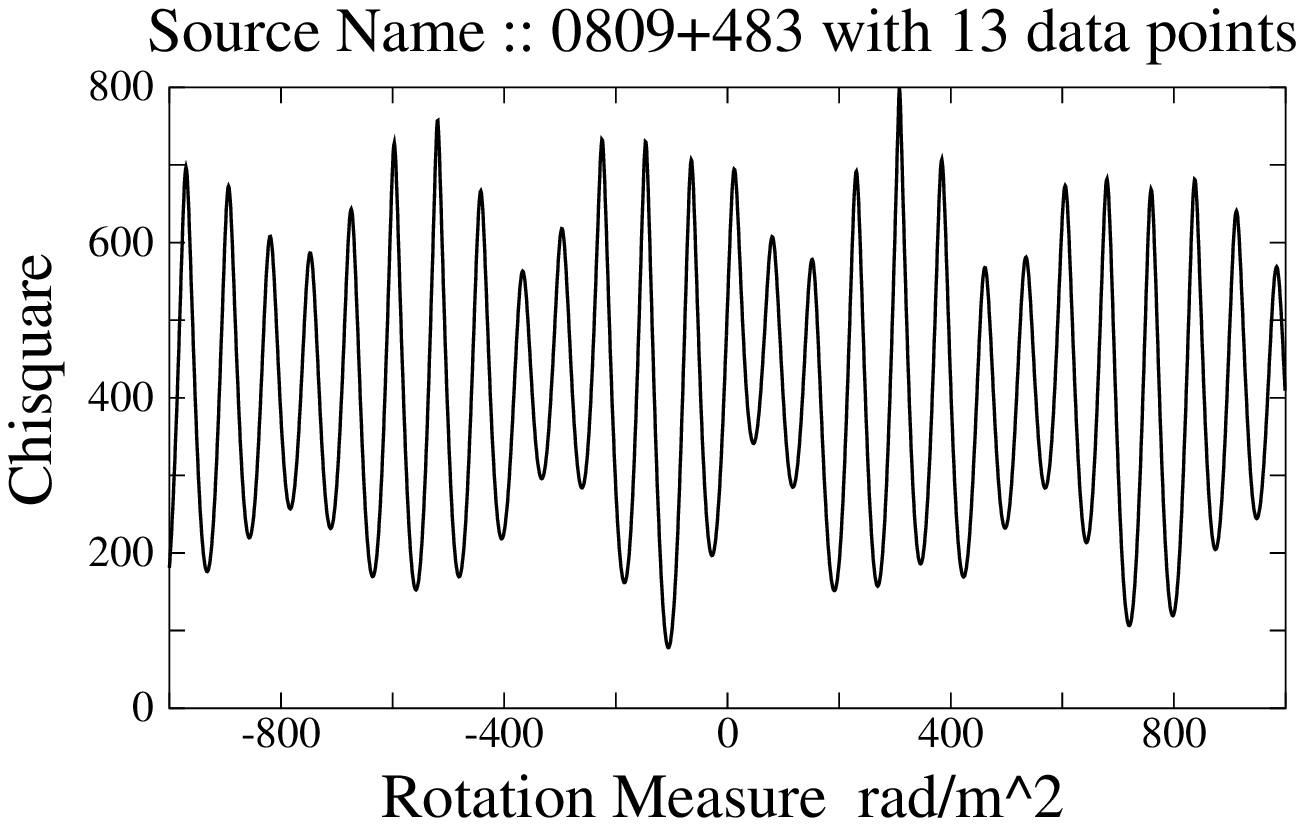}
\emi
}
\subf{
\bmi[t]{3in}
\includegraphics[scale=.55]{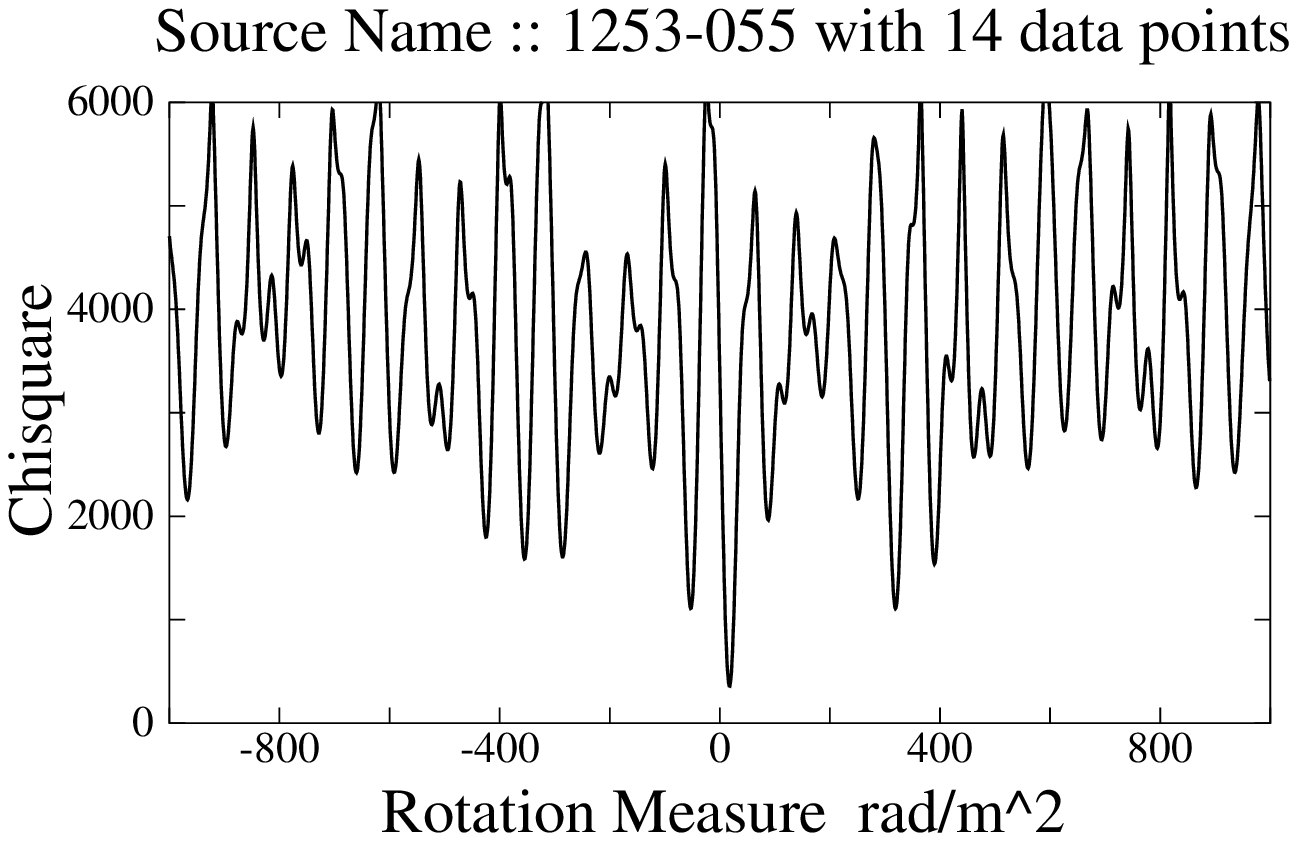}
\emi
}
\subf{
\bmi[b]{3in}
\includegraphics[scale=.55]{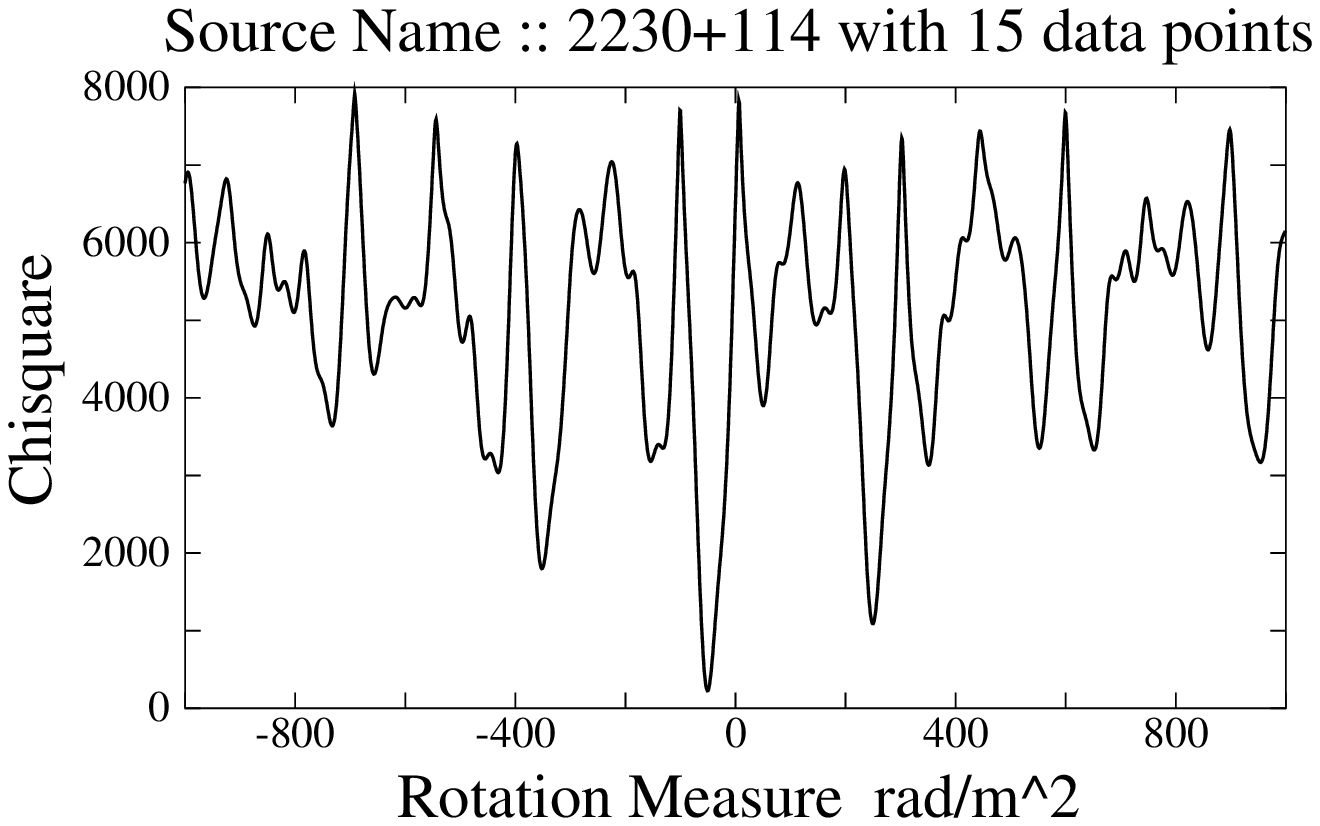}
\emi
}
\hspace{.9cm}
\subf{
\bmi[b]{3in}
\includegraphics[scale=.55]{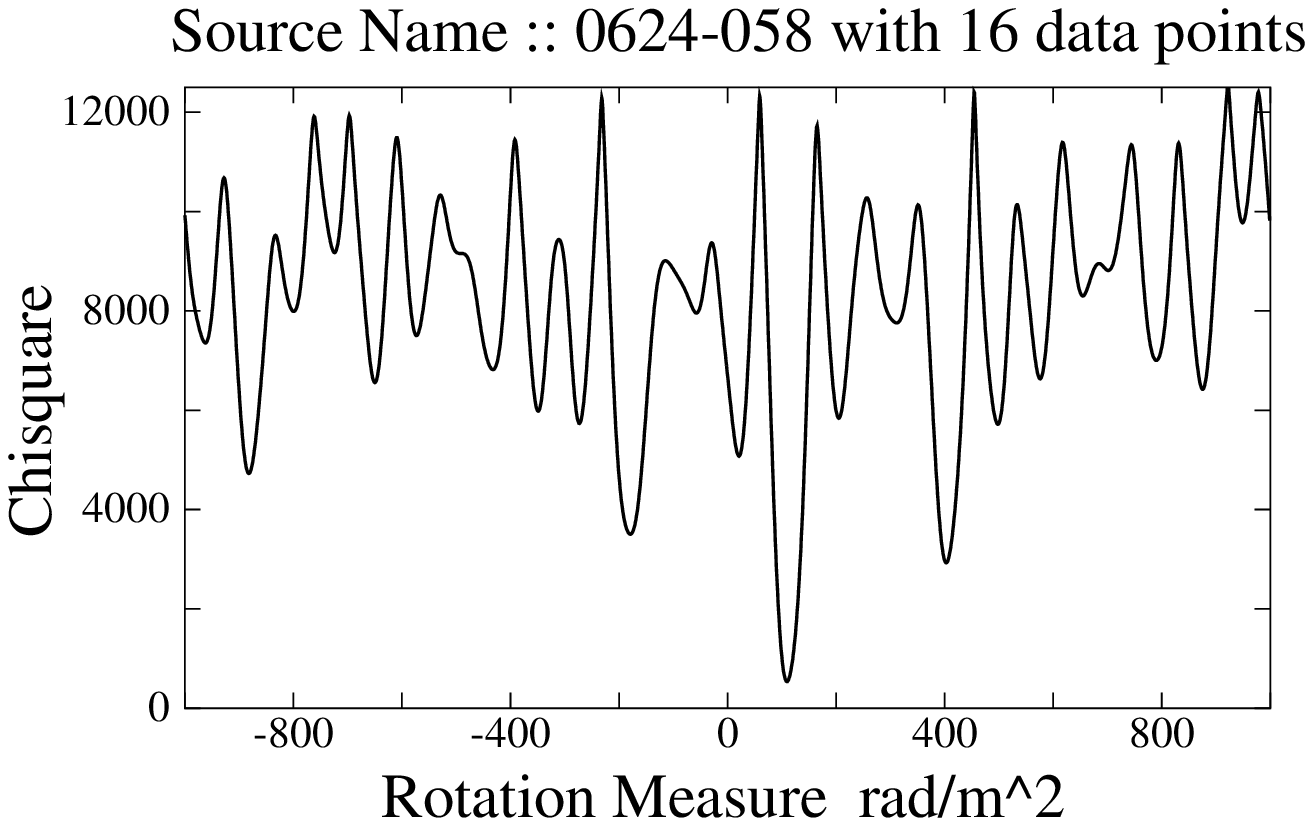}
\emi
}
\label{Chi2Cir}
\caption{ A representative sample of results showing
$\chi^2_{cir}$ as a function of $RM$. }
\efi

\section{Results and Discussion}
In Fig. \ref{Chi2Cir} we show $\chi^2_{cir}$ as a function 
of $RM$ for several representative cases. We find that $\chi^2_{cir}$ has 
several local minima. These minima corresponds to the standard ambiguity
in the determination of $RM$ due to $n \pi$ ambiguity in the polarization
angle. 
For large number $(>10)$ of data points, we find that unique minimum is
selected within the range $abs(RM) < 1000\ radians/m^2$ 
which turns out to be always in agreement with the value listed
in the Tabara and Inoue (1980). We see from Fig. \ref{Chi2Cir} that the
search interval in $RM$ does not have to be unduly small in order that the
program converge to the true minimum and hence does not require 
very long computational time. The procedure can be further optimized by
using several well known routines, such as simulated annealing, genetic
algorithms designed to find a global minimum efficiently. 
 
As we mentioned earlier Tabara and Inoue (1980) resolve the $n\pi$ 
ambiquity by searching the minimum in $\chi^2$ over a limited
 range of $RM$ which is selected on the basis of the position
of the source with respect to the galactic plane.
 We also use the same range of $RM$ in order to
 search for the global minima. 
In most cases we find that our results are in good agreement with
the ones given in Tabara and Inoue (1980), which have
been obtained using the standard linear fit. A randomly selected sample
of results using both the distributions (von Mises and the MB) 
 are shown in Table \ref{comparison}. The complete set of results
including the errors in $RM$ and $IPA$ can be found at the website
$http://home.iitk.ac.in/\tilde{\   } pkjain/RM\_DATA$.  

In many cases, however, we find that the 
global minima of $\chi^2_{cir}$, within the limited region of RM, is
not the same as the one obtained using standard $\chi^2$. 
The
global extrema for the log likelihood of MB distribution also differs
in some cases
from the extrema of both the $\chi^2_{cir}$ and $\chi^2$. We find
that for 16 sources the solution selected by $\chi^2_{cir}$ is
different from that selected by $\chi^2$. For all except two of these
cases the MB distribution prefers the solution obtained by using
$\chi^2_{cir}$. Overall the MB distribution gives results different
from $\chi^2_{cir}$ and $\chi^2$ for 30 and 44 sources respectively.  
  A sample
of results illustrating the different minima selected
by these different methods
are given in Table \ref{difference}. These differences arise since the
different procedures 
prefer 
different $n\pi$ combinations.
Hence in this sense the determination of rotation measure by
the standard $\chi^2$ method does have some bias, which is related
to the well known $n\pi$ ambiquity in the polarization angle. 

In Fig. \ref{comp1} 
we show $RM_L-RM_{MB}$ as a function of $RM_L$ where $RM_L$ is the rotation
measure 
calculated by minimization of $\chi^2$, i.e. the standard method, 
and $RM_{MB}$ is the result 
obtained by maximizing the log likelihood for the MB distribution. 
In figure \ref{comp1} we choose the solution $RM_{MB}$, among the different 
possible extrema of the log likelihood, such that $RM_{MB}$ is closest
to $RM_L$.  In Fig. \ref{comp2}, we show the comparison between the 
two determinations by choosing the absolute minima for the log likelihood
within the $RM$ search interval. 
 We also calculate the
correlation between $RM_L-RM_{MB}$ and $RM_L$ in order to investigate 
the existence of systematic difference between the two determinations. 
We find that if choose the log likelihood extrema such that it gives
$RM_{MB}$ closest to $RM_L$ then there does not exist any correlation
between $RM_L-RM_{MB}$ and $RM_L$. We find in this case that 
$n\rho^2 = 0.05$, where $\rho$ is the statistical correlation parameter. 
However if we choose the absolute minima for determination of $RM_{MB}$ 
we find that there does exist correlation between  $RM_L-RM_{MB}$ 
and $RM_L$ ($n\rho^2 = 26$). Hence in this sense these two determinations
of rotation measure show systematic difference, reflecting the existence
of bias due to $n\pi$ ambiquity.
Similar
calculation has been done for $IPA$ also. 
The results in this case for the correlation between the vectors
$[\sin(IPA_L),\cos(IPA_L)]$
and $[\sin(IPA_L-IPA_{MB}),\cos(IPA_L-IPA_{MB})]$
are shown in tables \ref{corr1} and \ref{corr2} for the two choices of
log likelihood extremum discussed above. 
It is clear from these tables that the $IPA$ determined by 
minimization of $\chi^2$ does show systematic difference from
that determined by maximizing log likelihood of the MB distribution.

\begin{table}
\begin{tabular} {|c|c|c|c|c|c|c|c|c|c|}\hline \hline
\multicolumn{1}{|c}{\bf Source} &
\multicolumn{3}{|c|}{\bf Circular fit ($MB$)} &
\multicolumn{3}{|c|}{\bf Circular fit (vM)} &
\multicolumn{3}{c|}{\bf Linear fit} \\ \hline 
 & $\mathbf{ RM}$ & $\mathbf {IPA}$ & 
 $\mathbf {\Delta(\ln {\cal L})\over dof}$ 
 & $\mathbf{ RM}$ & $\mathbf {IPA}$ & 
 $\mathbf {\chi^2_{cir}\over dof}$ 
 & $\mathbf{ RM}$  & $\mathbf {IPA}$ & 
 $\mathbf {\chi^2 \over dof}$ \\ \hline

0002+125 &-14.3 &177.7 &2.6 & -15.9 & 178.3 & 3.6 & -16.0 &  178.0 & 3.7 \\ \hline  
0010+005 &-10.5 &118.0 & 2.6 & -10.7 & 117.7 & 1.0 & -10.7 &  117.0 & 1.0 \\ \hline  
0035-024 &5.1 &164.8 &2.3 & 7.4 &  164.6 & 2.0 &  7.7 &  164.0 & 2.1 \\ \hline  
0128+061 & 3.8 & 96.0 & 1.6 & 5.1 &  95.7 &  2.0 &  5.1 &  95.0 &  2.1 \\ \hline  
0245-558 &22.7 & 21.5 & 0.9 & 22.0 &  21.8 & 0.3 & 22.0 & 21.8 &  0.3 \\ \hline  
0422+004 & -29.9 &24.2 &4.3 &  -11.6 & 19.7 & 5.0 & -8.1 & 18.0 &  5.2 \\ \hline  
0538+498 & 308.4 &96.2 & 2.3 &  310.7 & 96.3 & 1.8 & 311.3 & 94.4 & 2.2 \\ \hline  
0540+187 & 31.5 & 96.4 &2.5 &31.0 & 100.0 & 1.4 & 31.0 & 100.0 & 1.4 \\ \hline  
0547-408 & 49.0 & 110.9 &1.7 & 47.9 & 111.0 & 0.9 & 47.9 & 111.0 & 0.9 \\ \hline  
0735+178 &-299.7 &90.1 &4.1 & -300.8 & 86.8 & 12.7 & -300.9 & 85.6 & 14.3 \\ \hline  
0859-140 &10.1 & 75.7 &2.3 & 8.3 & 84.9 & 3.9 &  8.3 & 84.9 & 3.9 \\ \hline  
1030+585 &2.7 &103.4 &4.0 & 2.7 & 103.1 & 9.2 &  1.7 & 105.5 & 10.5 \\ \hline  
1203+645 &109.9  &69.9 &2.0 & 111.6 & 60.0 & 1.9 & 111.9 & 57.9 & 2.0 \\ \hline  
1319+428 &-13.5 &70.2 &3.0 & -15.2 & 74.9 & 1.9 & -15.2 & 75.0 & 1.9 \\ \hline  
1424-418 &-199.2 &82.0 &1.7 & -199.5 & 82.0 & 0.6 & -199.5 & 82.0 & 0.6 \\ \hline  
1755-162 &129.3 &106.1 &3.0 &  129.9 & 107.3 & 5.7 & 129.9 & 107.3 & 5.7 \\ \hline  
2210+016 & -68.4& 146.8 &1.8 & -67.3 & 139.5 & 0.9 & -67.3 & 139.3 & 0.8 \\ \hline  
2251+158 &-55.7 & 21.3&2.9 & -55.7 & 24.1 & 6.4 & -55.7 & 24.0 & 6.4 \\ \hline  
2348+643 &-950.2 &143.1 &1.2 & -953.8 & 151.7 & 0.5 & -953.9 & 151.0 & 0.5 \\ \hline  
2356-611 &20.8 &14.8 &2.7 &  19.6 & 17.5 & 2.7 & 19.6 & 17.5 & 2.7 \\ \hline  
\hline  
\end{tabular}
\caption{ A randomly selected sample of the results using the
likelihood analysis (Circular fit) with the MacDonald-Bunimovitch (MB) 
and von Mises (vM) 
distributions  
and its comparison with the standard linear fit. The $RM$
values are given in units of $rad/m^2$ and the $IPA$ are in degrees.
The $\Delta(\ln {\cal L})/ dof$, $\chi^2_{cir}/dof$ and $\chi^2/dof$
are the measures of goodness of fit for the three different cases, dof
stands for degrees of freedom.
}
\label{comparison}
\end{table}

\begin{table}
\begin{tabular}{|c|c|c|c|c|}
\hline
Source &  data &  linear fit&circular fit (vM) & circular fit (MB)  \\
 &pts &	$(RM,IPA,\chi^2)$ &	$(RM,IPA,\chi^2_{cir})$ & $(RM,IPA,
\Delta[\ln{\cal L}])$ \\
\hline
  0127+233 & 4 &$(-147.2, 20.5, 6.5)$ &  $(-147.1, 20.0, 5.9)$ &$(-147.0, 18.5, 5.3)$  \\
& &	$(-71.5, 3.4, 9.2)$  &  $(-72.3, 3.1, 5.8)$ & $(-72.7, 3.1, 4.4)$\\
\hline
0506-612 & 4 & $(-183.9,22.6,31.6)$ & $(-183.1,21.9,29.2)$ & $(-196.6,23.9,9.5)$ \\
    & &  $(116.1,150.3,33.2)$ & $(121.6,147.3,28.6)$ &$(133.9,140.6,6.9)$  \\ 
\hline
0758+143  & 8 & $(128.0,99.1,15.9)$ & $(128.5,99.2,14.5)$ & $(128.5,104.9,17.6)$ \\
 & &      $(-112.1,3.7,16.5)$ & $(-115.9,16.0,13.0)$ & $(-114.2,11.5,13.1)$ \\
\hline
0806-103 & 6 & $(334.8,17.4,48.9)$ & $(335.0,16.7,46.3) $ & $(337.4,5.0,17.3) $ \\
    & & $(259.2,54.4,53.6)$ & $(261.0, 55.3,40.7)$ &  $(263.5, 56.2,10.0)$\\
\hline
1335-061 & 6 & $(-93.9,32.7,24.7)$ & $(-94.1,32.9,23.5)$ & $(-87.9,34.4,14.7)$ \\
   & & $(-59.0,17.0,27.2)$ & $(-60.8,17.6,19.9)$ & $(-63.9,18.9,11.5)$ \\
\hline
1434+036 & 5 & $(-108.3,46.7,6.0)$ & $(-108.3,46.5,5.8)$ & $(-108.6,43.0,7.3)$ \\
	& & $(-76.1,35.2,7.0)$ & $(-75.1,35.2,4.7)$ & $(-74.8,35.1,6.0)$ \\
\hline
1957+405 & 6 & $(34.2,141.0,5.7)$ &$(53.0,135.9,4.9)$ & $(55.1,135.5,6.3)$\\
    & & $(81.5,128.2,5.9)$ & & \\
\hline
2120+168 & 5 & $(-115.6,12.7,18.0)$ & $(-115.9,13.2,17.5)$ & $(-114.5,178.5,12.5)$ \\
    & & $(-37.1,156.0,19.8)$ & $(-37.2,154.8,16.8)$ &$(-40.0,162.3,9.9)$ \\ 
\hline
\end{tabular}
\caption{A sample of sources for which the global minima of 
$\chi^2_{cir}$ and $\chi^2$ occur at different RM values. 
A single value for the circular fit for the source 1957+405 indicates that
only one extrema was found.}
\label{difference}
\end{table}

\begin{table}
\hskip 1.0in
\begin{tabular}{|c|c|c|}
\hline
& $\sin(IPA_L-IPA_{MB})$ & $\cos(IPA_L-IPA_{MB})$ \\
\hline
$\sin(IPA_L)$ & 2.4 & 4.2\\ 
\hline
$\cos(IPA_L)$ & 10.1 & 11.8\\ 
\hline
\end{tabular}
\caption{Correlation matrix $n\rho^2_{A,B}$ between $[\sin(IPA_L),\cos(IPA_L)]$
and $[\sin(IPA_L-IPA_{MB}),\cos(IPA_L-IPA_{MB})]$ for the case when the
extremum of log likelihood of  MB distribution is chosen such that it 
gives rotation measure nearest to that obtained by minimizing $\chi^2$.
$IPA_L$ and $IPA_{MB}$ are the estimates of the intercept 
of the fit determined using the 
linear method and the likelihood analysis using MB distribution 
respectively.
 }
\label{corr1}
\end{table} 

\begin{table}
\hskip 1.0in
\begin{tabular}{|c|c|c|}
\hline
& $\sin(IPA_L-IPA_{MB})$ & $\cos(IPA_L-IPA_{MB})$ \\
\hline
$\sin(IPA_L)$ & 6.2 & 27.3\\ 
\hline
$\cos(IPA_L)$ & 5.1 & 10.1\\ 
\hline
\end{tabular}
\caption{Correlation matrix $n\rho^2_{A,B}$ between $[\sin(IPA_L),\cos(IPA_L)]$
and $[\sin(IPA_L-IPA_{MB}),\cos(IPA_L-IPA_{MB})]$ for the case when the
absolute extremum of log likelihood of MB distribution is chosen within
the search range of the rotation measure.
$IPA_L$ and $IPA_{MB}$ are the estimates of the intercept 
of the fit determined using the 
linear method and the likelihood analysis using MB distribution 
respectively.
}
\label{corr2}
\end{table} 

It has been found that there are few cases shown in Fig. \ref{LargeChi}  
where the $\chi^2_{cir}$ per 
degree of freedom is very large $(>30)$. We have verified that the
corresponding fit using MB distribution is also quite close to
the fit using $\chi^2_{cir}$. In the graphs we prefer to show
the results obtained using vM distribution because the corresponding 
statistic $\chi^2_{cir}$ has 
a more direct interpretation than the likelihood difference 
$\Delta(\ln{\cal L})$ in the 
case of MB distribution. For the cases displayed in
figure our determination of $RM$ is quite close 
to Tabara and Inoue (1980) values. We have verified that
the standard linear $\chi^2$ is also very large for these cases,
as expected. Hence
it seems that the straight line is not a good fit for these sources.
The dominant reason for this behavior appears to be that different 
regions in the source, which in general have different polarization, 
may dominate at different wavelengths.  
In order to make an unbiased extraction of $RM$ for these sources it
may be necessary to use a limited range of wavelength 
(Vall\'{e}e 1980) chosen on the basis of some physical characteristics
of the source.
Alternatively one can introduce higher order terms in $\lambda^2$
in the fit. The procedure proposed in this paper can be easily 
extended to include such terms. 

\bfi
\subf{
\bmi[t]{3in}
\includegraphics[scale=1.0]{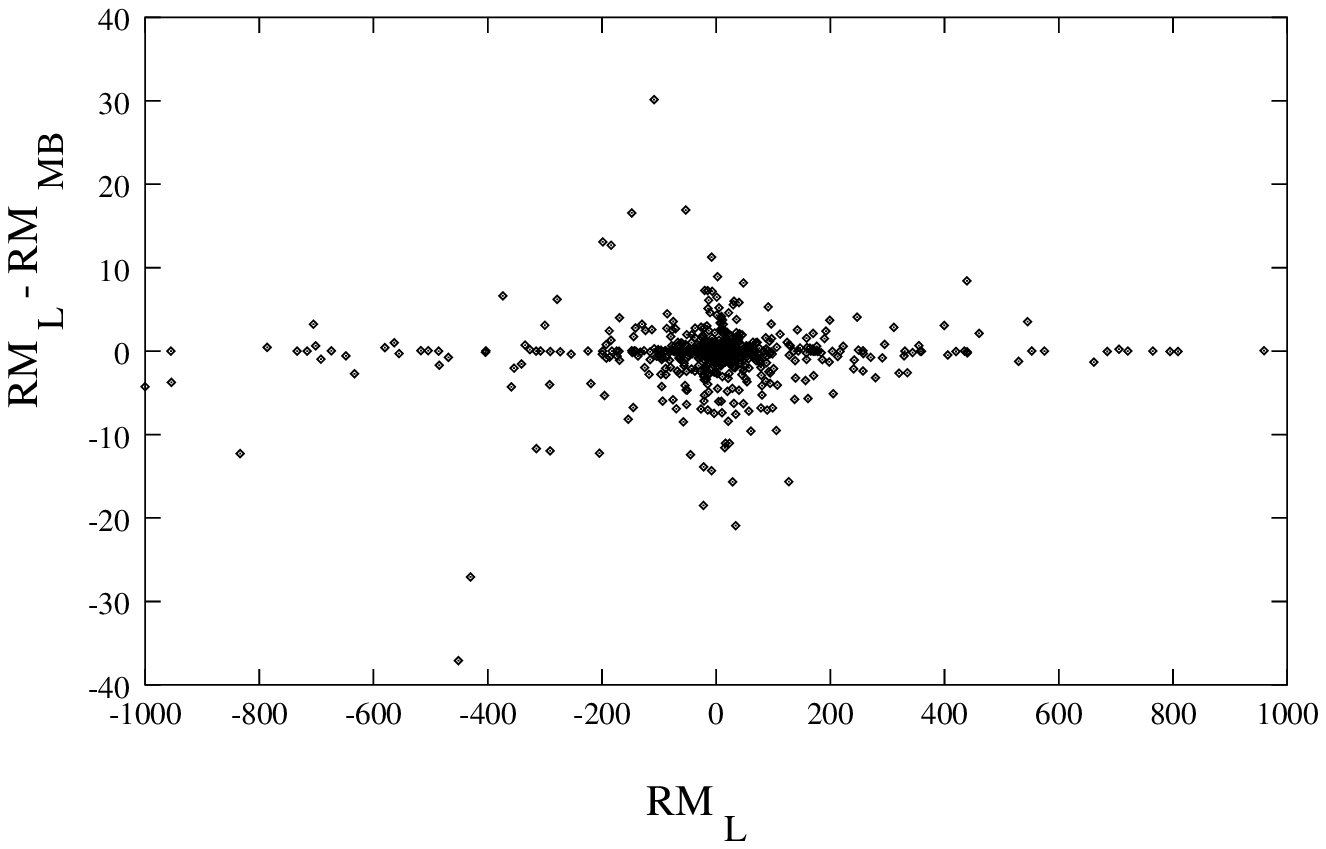}
\emi
}
\caption{Comparison of the $RM$ determined by minimization of 
$\chi^2$ $(RM_L)$ with that determined by use of the MB distribution
$(RM_{MB})$. The units of $RM$ are $radian/m^2$. 
Here the $RM_{MB}$ solution is selected such that it
gives results closest to the linear determination, even if the absolute
maxima in the log likelihood occurs at a different location. The range in
y-axis has been limited to $\pm 40$ for the sake of clarity. There were
three data points outside this range which are not shown in the graph.}
\label{comp1}
\efi 

\bfi
\subf{
\bmi[t]{3in}
\includegraphics[scale=1.0]{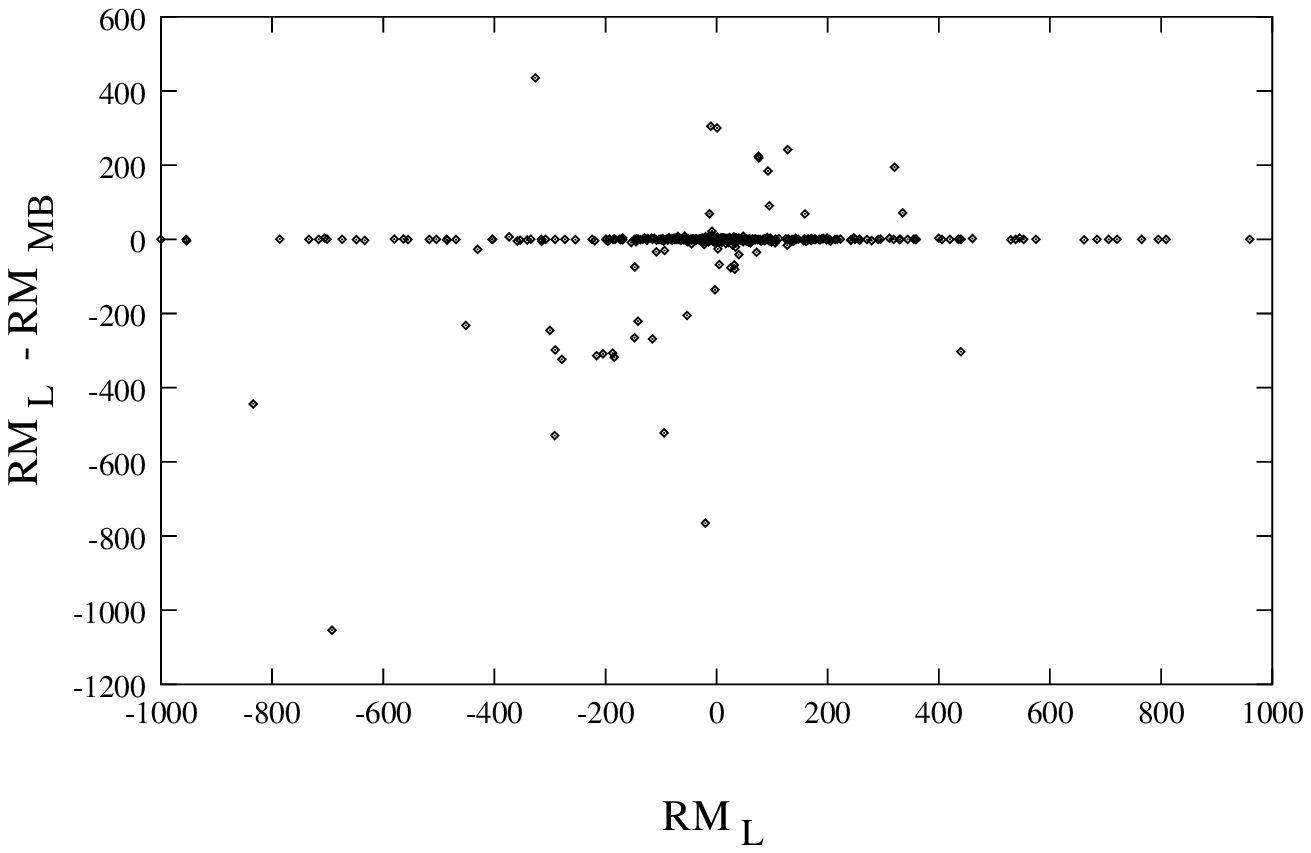}
\emi
}
\caption{Comparison of the $RM$ determined by minimization of 
$\chi^2$ $(RM_L)$ with that determined by use of the MB distribution
$(RM_{MB})$. The units of $RM$ are $radian/m^2$. 
Here the $RM_{MB}$ solution is selected 
by the absolute
maxima of the log likelihood. 
}
\label{comp2}
\efi 

\section{Conclusion}
We have proposed a new method for the computation of rotation measures
from spectral polarization data. The method takes into account
the circular nature of polarization observable and 
is based on the maximum likelihood analysis. It employs the MB distribution
of polarization angle which can be derived by assuming that the electric field 
vector follows a normal distribution. Since the measured polarization
angles are expected to follow this distribution the method will give an 
 efficient and unbiased estimate of $RM$ and $IPA$. 
A simpler procedure for determination
of rotation measure, based on the von Mises distribution, is also 
presented. Due to its simplicity and invariance under circular 
coordinate transformation, the von Mises distribution turns out to
very useful for the purpose of providing an initial ansatz which can 
then be used for the more accurate $RM$ determination using the MB
distribution. For many purpose this initial ansatz may itself be
sufficiently accurate. The von Mises distribution leads us to 
define a new statistic, $\chi^2_{cir}$, which turns out to be very useful for 
handling circular data. In particular it provides us with an easy method
to search over the RM range to find the absolute minima without 
requiring us to look at all possible $n\pi$ combinations which can
be added to the polarization angle. For this reason it turns out to
be computationally  very fast compared to the standard
$\chi^2$ if the data set is very large $(N>10)$. Since large data
sets are required in order to obtain an unambiquous value of RM,
the present method can be very useful.
We further find that due to the $n\pi$ ambiquity the rotation
measure determined by the standard $\chi^2$ method shows systematic
difference from the procedure using the MB distribution. This difference
dissapears if we insist that the same $n\pi$ combination is used
for the two determinations. However we find that systematic difference
in the determination of $IPA$ remains even if we require that the two
methods use the same $n\pi$ combination.  
  The procedure
can also be easily extended to extract higher order terms, i.e. nonlinear
dependence on wavelength squared, which often arise due to source 
morphology and can play an important role in characterizing the source.
 Hence our procedure may
provide us with a useful tool for analysing large polarization data sets,  
if they become available in future, for the purpose of extracting 
rotation measures and other information about polarizations from 
radio sources.

\section{Acknowledgements:} We thank John Ralston and G. K. Shukla
for very useful comments. Financial assistance for this 
work was provided by a grant from Department of Science and
Technology.
 
\vspace{3cm}
\center{\Large \bf References}
\begin{itemize}
\item[] Batschelet, E. 1981, {\it Circular Statistics in Biology},
(London: Academic Press)
\item[] Born, M. and Wolf E. 1980, {\it Principles of Optics}, Pergamon
Press.
\item[] Brosseau, C. 1998, {\it Fundamentals of Polarized Light, A
Statistical Optics Approach}, John Wiley \& Sons, Inc.
\item[] Broten, N. W., MacLeod, J. M. and Vall\'{e}e, J. P. 1988,
{\it Astrophysics and Space Science} {\bf 141}, 303
\item[] Bunimovitch, V. I. 1949, {\it Zhur. Tekh. Fiz.} {\bf 19},
1231.
\item[] Cramer, H. 1958, {\it Mathematical Methods of Statistics},
Princeton Univ. Press, New Jersey.
\item[] Fisher, N. I. 1993, {\it Statistics of Circular Data},
(Cambridge)
\item[] Press, W. H. {\it et al} 1986, {\it Numerical Recipes}, Cambridge
University Press.
\item[] Freund, J. E. 1992, {\it Mathematical Statistics}, Prentice-Hall Inc.
\item[] Jain, P. and Ralston, J. P. 1999,  {\it Mod. Phys. Lett.}
{\bf A14}, 417
\item[]Kendall, D. G. and Young, A. G. 1984, {\it Mon. Not. R. Astron. Soc}
{\bf 207}, 637.
\item[] MacDonald, D. K. C. 1949, {\it Proc. Camb. Phil. Soc.} {\bf 45},
368.
\item[]  Mardia, K. V. 1972, {\it Statistics of Directional
Data},
(London: Academic Press)
\item[] Ralston, J. P. and Jain, P. 1999, {\it International
Journal of Modern Physics} {\bf D 8}, 537.
\item[] Simard-Normandin, M., Kronberg, P. P. and Button, S. 1981,
{\it The Astrophysical Journal Supplement Series}, {\bf 45}, 97
\item[] Tabara, H. and  Inoue, M. 1980, 
{\it Astron. and Astrophys, Suppl.
Ser.} {\bf 39}, 379.
\item[] Vall\'{e}e, J. P. 1980, {\it Astron. Astrophys.} {\bf 86}, 251.  
\item[] Vall\'{e}e, J. P. 1997, {\it Fundamentals of Cosmic Physics} 
{\bf 19}, 1.  
\item[] Zeldovich Ya. B., Ruzmaikin A. A. and Sokoloff D. D. 1983,
{\it Magnetic fields in Astrophysics}, 
(Gordon and Breach Science Publishers).
\end{itemize}

\newpage
\bfi
\hspace{1.5cm}
\bmi[b]{6in}
\includegraphics[]{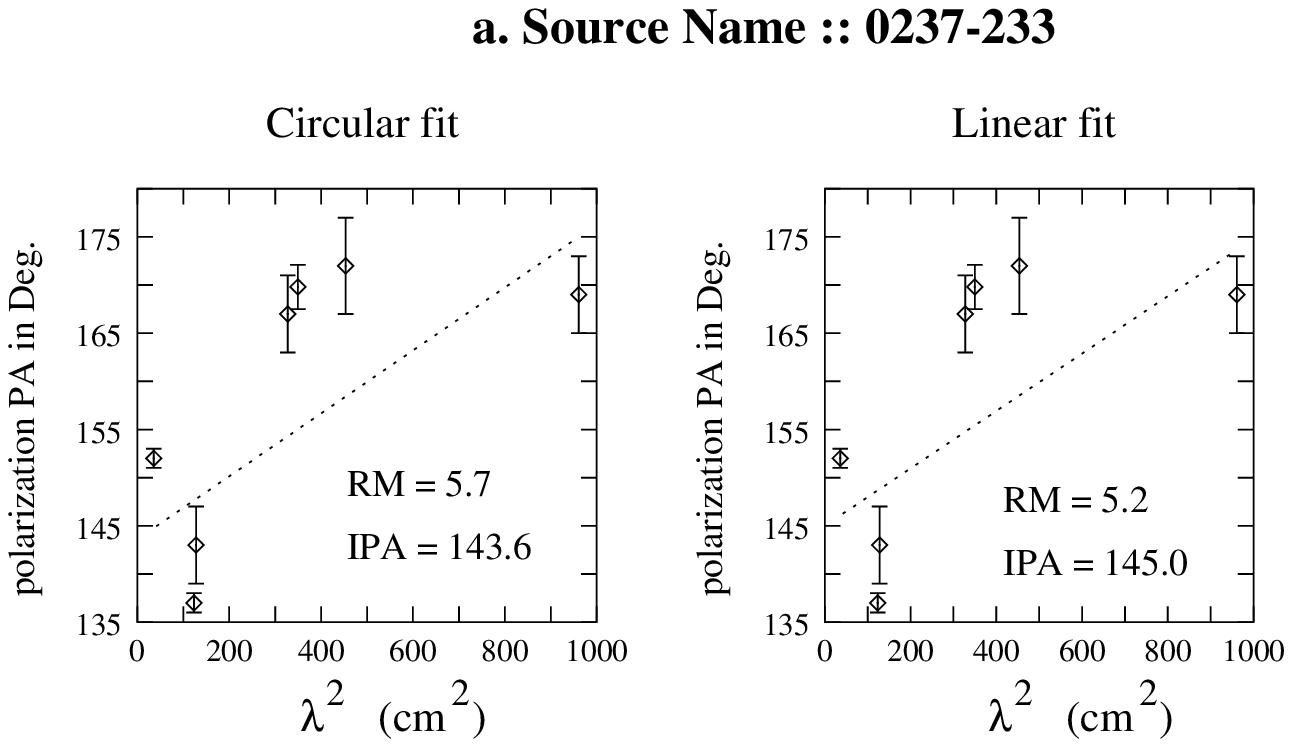}
\emi
\efi

\bfi
\hspace{1.5cm}
\bmi[t]{6in}
\includegraphics[]{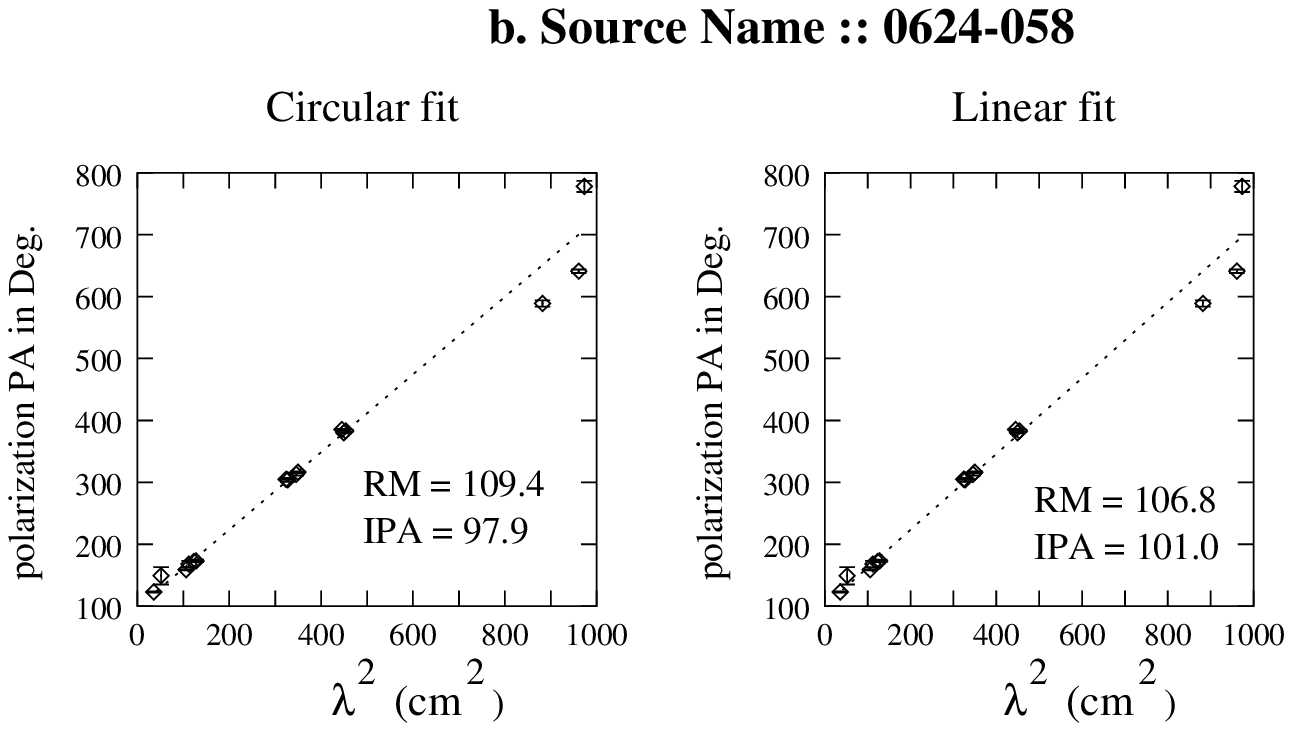}
\emi
\efi

\bfi
\hspace{1.5cm}
\bmi[b]{6in}
\includegraphics[]{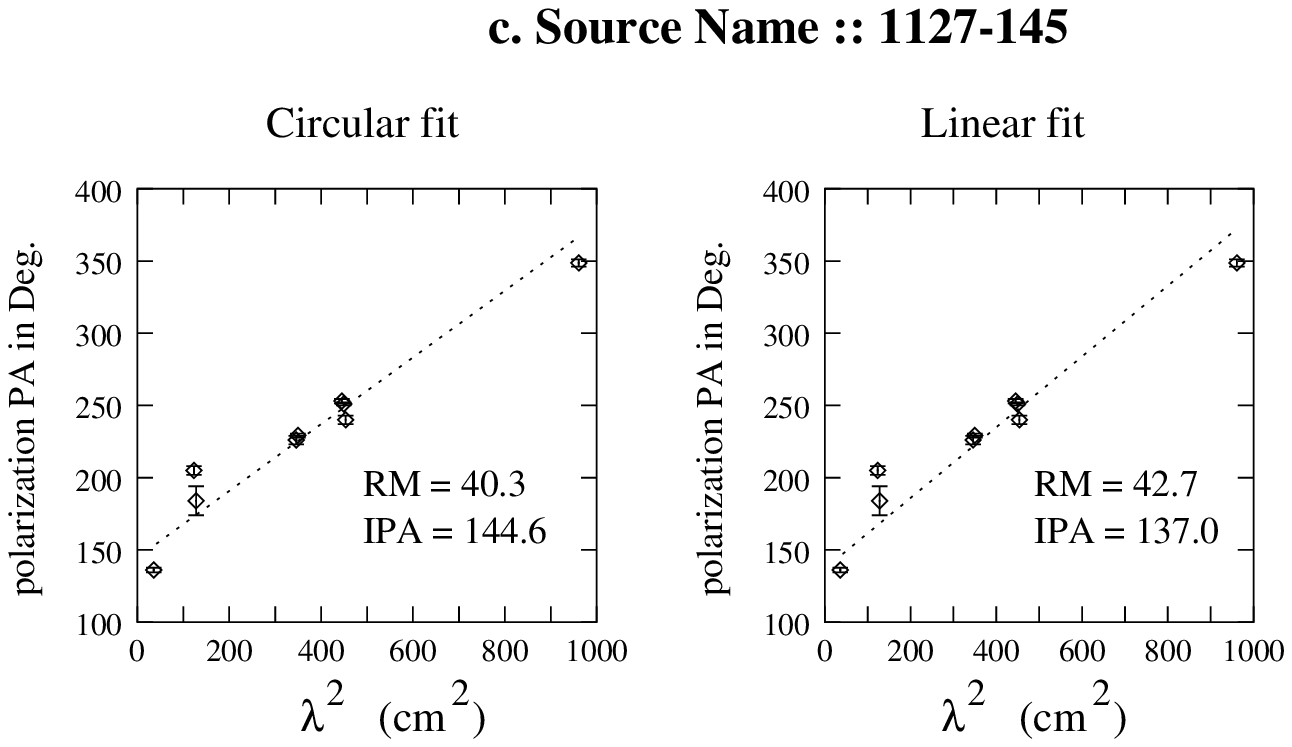}
\emi
\efi

\bfi
\hspace{1.5cm}
\bmi[t]{6.0in}
\includegraphics{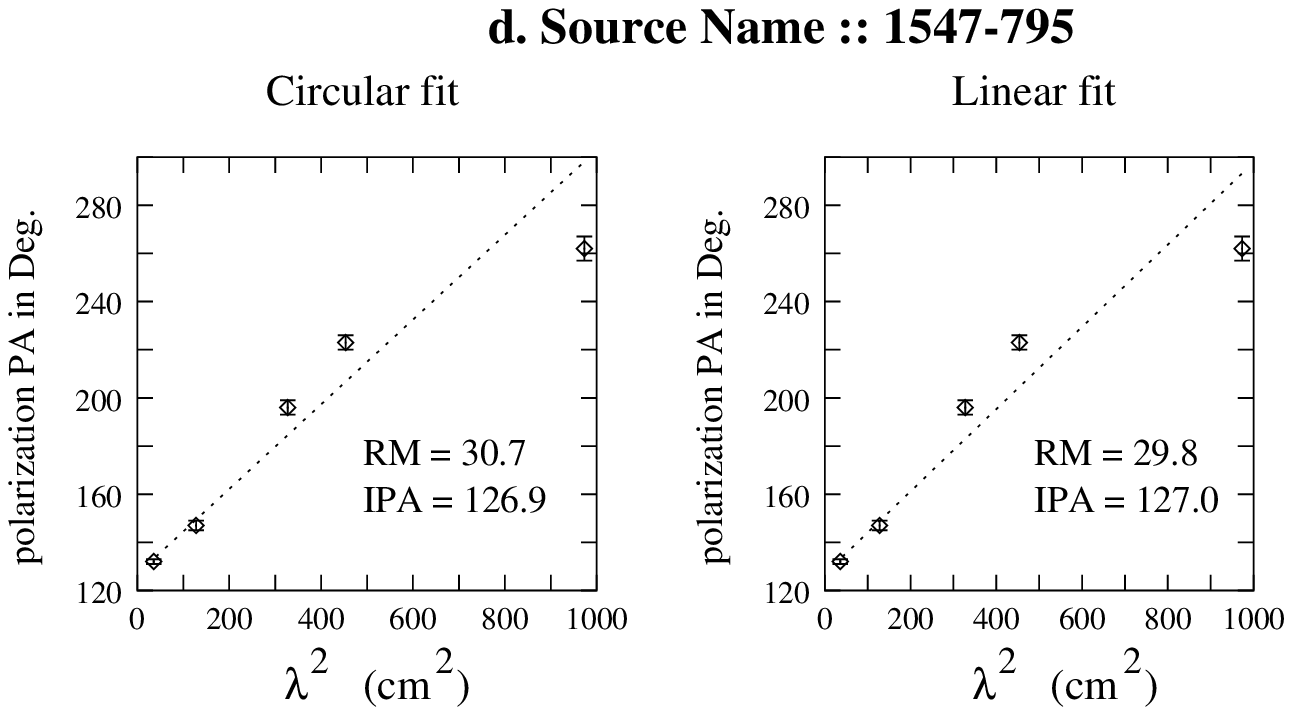}
\emi
\efi

\bfi
\hspace{1.5cm}
\bmi[b]{6.0in}
\includegraphics{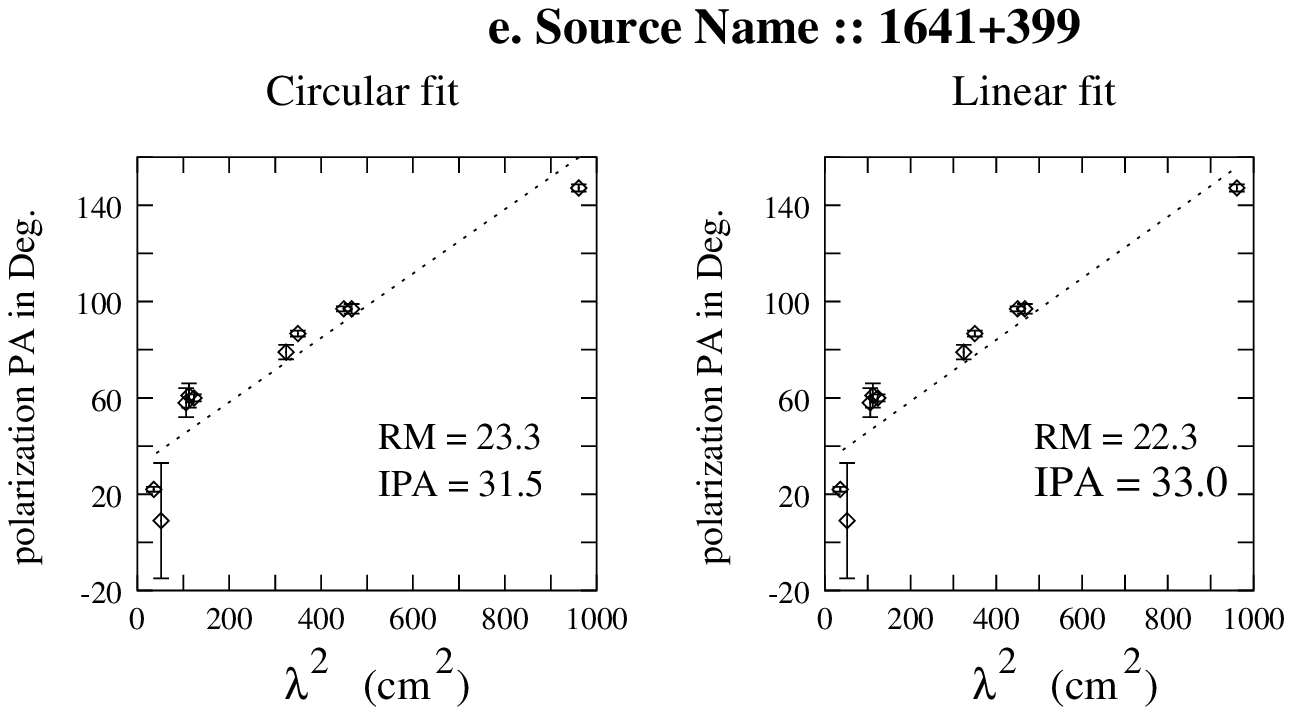}
\emi
\efi

\bfi
\hspace{1.5cm}
\bmi[t]{6.0in}
\includegraphics{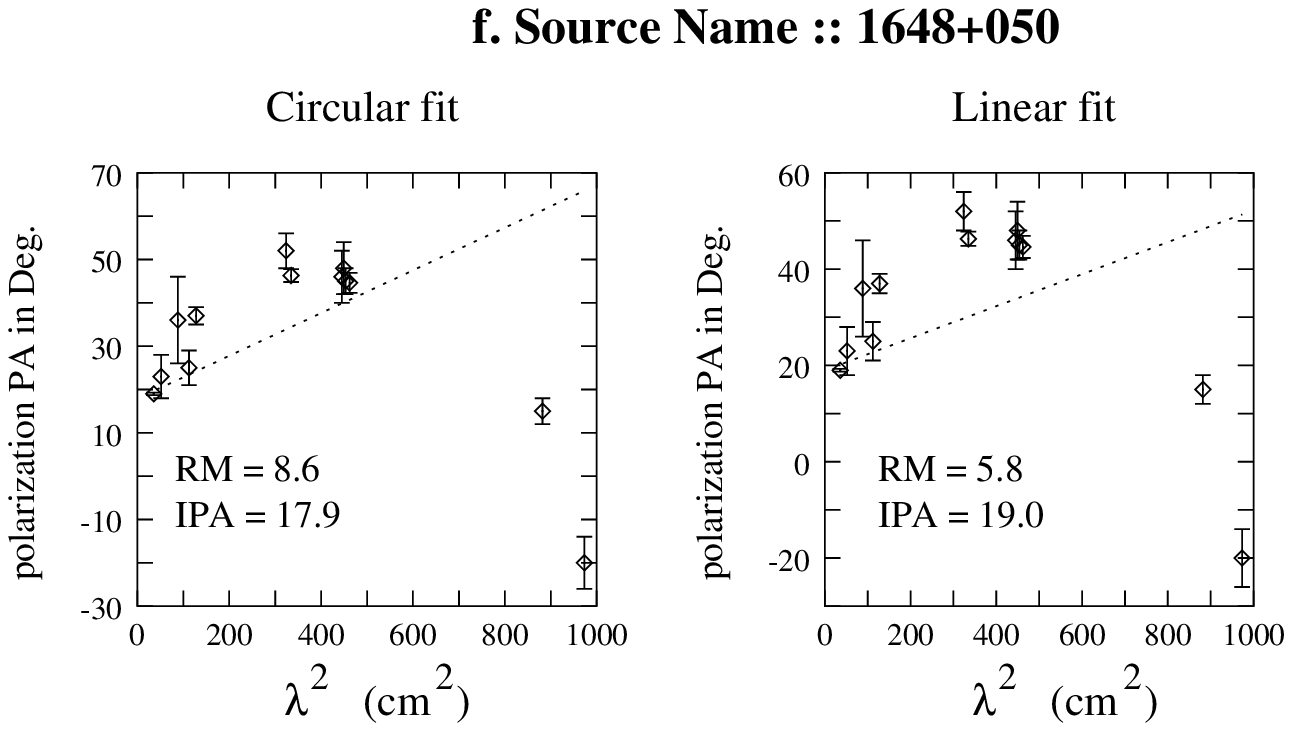}
\emi
\caption{(a) to (f) :: Comparison of the fit found for  
the cases where the $\chi^2_{cir}/dof$ is very large $(>30)$. In all 
the graphs the units of $RM$ and $IPA$
are $rad/m^2$ and Degrees respectively. }
\label{LargeChi}
\efi

\end{document}